\documentclass[twocolumn,showpacs,preprintnumbers]{revtex4-2}	
\usepackage[caption=false]{subfig}
\usepackage{amsmath}  
\usepackage{amsfonts}  
\usepackage{graphicx} 
\usepackage{enumerate}
\usepackage{tikz}
\usetikzlibrary{shapes,arrows, arrows.meta, calc, positioning}
\usepackage[colorlinks=true,linkcolor=blue,urlcolor=blue,citecolor=blue]{hyperref}
\usepackage{xcolor}
\begin{document}
\newcommand{\ds}{\displaystyle}
\newcommand{\be}{\begin{equation}}
\newcommand{\en}{\end{equation}}
\newcommand{\bea}{\begin{eqnarray}}
\newcommand{\ena}{\end{eqnarray}}

\title{Can wormhole spacetimes in Unimodular Gravity be supported by ordinary matter? A general proof of the exotic matter requirement}
\author{Mauricio Cataldo}
\altaffiliation{mcataldo@ubiobio.cl} 
\affiliation{Departamento de
F\'\i sica, Universidad del Bío-B\'io, Casilla 5-C,
Concepci\'on, Chile.\\}
\affiliation{Centro de Ciencias Exactas, Universidad del Bío-Bío, Casilla 447, Chillán, Chile. \\}
\author{Norman Cruz}
\altaffiliation{norman.cruz@usach.cl} 
\affiliation{Departamento de Física, Universidad de Santiago de Chile, Avenida Ecuador 3493, Santiago, Chile. \\}
\affiliation{Center for Interdisciplinary Research in Astrophysics and Space Exploration (CIRAS), Universidad de Santiago de Chile, Av. Libertador Bernardo O'Higgins 3363, Estación Central, Chile. \\}
\author{Patricio Salgado}
\altaffiliation{patsalgado@unap.cl} 
\affiliation{Facultad de Ciencias, Univesidad Arturo Prat, Avda. Arturo Prat 2120, 
Iquique, Chile. \\}
\affiliation{Instituto de Ciencias Exactas y Naturales (ICEN), Univesidad Arturo Prat, Avda. Playa Brava 3256, Iquique, Chile.\\}

\date{\today} 
\begin{abstract}
We establish a general no--go theorem demonstrating that all traversable wormhole configurations in Unimodular Gravity necessarily require exotic matter. The proof relies solely on the geometric flaring-out condition, $b'(r_0) \leq 1$, which directly implies that $\rho(r_0) + p_r(r_0) \leq 0$ at the throat. This condition represents a violation of the Null Energy Condition and, consequently, of the Weak and Strong Energy Conditions, independently of the particular choice of shape function, redshift function, or equation of state. This result holds for both tidal and zero-tidal-force configurations, showing that the requirement of exotic matter is a fundamental geometric consequence of the traversability condition rather than an artifact of specific solution choices. Therefore, Unimodular Gravity shares this fundamental constraint with General Relativity.


\vspace{0.5cm}
\end{abstract}
\smallskip
\maketitle 


\section{Introduction}
Wormholes represent solutions to Einstein's field equations that act as tunnels or bridges connecting two regions of spacetime, thus providing a shortcut between them. These geometries were explored by Kip Thorne and his collaborators, who established the conditions for traversability~\cite{Morris:1988cz1,Morris:1988cz2,Visser}. Unlike black holes, which possess a unidirectional event horizon, traversable wormholes lack horizons and theoretically allow bidirectional passage of matter. Their geometry is characterized by a throat connecting two mouths that may exist in different regions of the same universe or potentially in separate universes.

Since Morris and Thorne introduced the concept of traversable wormholes and demonstrated that exotic matter was required for spherically symmetric wormholes to be traversable, multiple attempts have been made to obtain geometric configurations that would allow for traversable wormhole solutions supported by ordinary matter. These attempts were conducted within the framework of General Relativity (GR) or Brans-Dicke gravity~\cite{BD,BD1}, considering symmetries other than spherical, altering asymptotic flatness, and incorporating time dependencies to study the evolution of these wormholes.

This fundamental obstacle persisted, and the necessity of exotic matter violating the null energy condition (NEC) to keep the wormhole throat open could not be avoided. In 1998, Hochberg and Visser reinforced this requirement, demonstrating that the violations of the NEC associated with traversable wormholes are completely generic. The NEC violations will occur at or near the throat of any spacetime configuration that deserves to be called a traversable wormhole. This result is completely general and independent of symmetry, asymptotic flatness, or time dependence~\cite{Visser y Hochberg}.

In the last two decades, various theoretical approaches have attempted to overcome this limitation. Modified gravity theories have shown promising results, demonstrating that higher-order curvature terms can act as a gravitational fluid that sustains wormhole geometries while ordinary matter satisfies all energy conditions~\cite{Lobo}.

Among these theoretical frameworks, Einstein-Cartan theory~\cite{EC}, which incorporates torsion, has provided exact solutions with spherical symmetry using scalar fields that satisfy the weak energy condition, where torsion plays a crucial role in stabilization. The Randall-Sundrum brane-world model provides another viable approach, where matter on the brane may be non-exotic while extra-dimensional effects generate the necessary geometry~\cite{RS}.

More recently, the Einstein-Dirac-Maxwell approach proposed by Blázquez-Salcedo and collaborators~\cite{BS} using massive fermions in a singlet spin state generated great interest, although it was subsequently questioned due to differentiability problems at the throat~\cite{CC,BBKS}.

These theoretical advances suggest that, although exotic matter remains necessary within the framework of classical GR, extensions of this theory offer promising paths for constructing physically plausible wormholes using only ordinary matter, where exotic effects are provided by the modified geometric structure of spacetime.

Unimodular gravity is one of the oldest and most extensively studied 
alternatives to GR. Its origin can be traced back to Einstein's 1919 
trace-free field equations~\cite{Alvarez,Smolin,Fernandez}, and it has received 
renewed attention over the past decade from several different perspectives. Its defining 
feature is the imposition of a fundamental constraint: the determinant of the spacetime 
metric is required to remain fixed throughout the manifold, $\sqrt{-g} = \epsilon_0$. 
This constraint reduces the full diffeomorphism invariance of GR to volume-preserving 
diffeomorphisms and leads to trace-free field equations in which the cosmological 
constant does not appear as a free parameter of the action. Instead, the divergence of 
the traceless field equations yields the relation
\begin{equation}
\frac{R_{;\nu}}{4} = 8\pi G\left(T^{\mu\nu}_{;\mu} - \frac{T_{;\nu}}{4}\right),
\end{equation}
which in general allows for a varying $\Lambda$~\cite{Fernandez,Fabris2023}. Only if the 
conservation of the energy-momentum tensor $T^{\mu\nu}_{;\mu} = 0$ is additionally 
imposed does the cosmological constant reduce to a constant of integration, and the 
full Einstein equations with a fixed $\Lambda$ are recovered~\cite{Smolin,Fabris2023}. 
This structure implies that Unimodular Gravity and GR, while sharing the same classical field equations 
under this additional assumption, are known to differ at the quantum and perturbative 
levels~\cite{Fernandez,Bufalo2015,Fabris2023}. At the foundational level, Alencar~\cite{Alencar2026} 
has recently argued that the unimodular condition $\sqrt{-g} = \epsilon_0$ may carry 
deeper geometric significance, suggesting that this constraint need not be regarded as 
a mere gauge choice but may instead reflect a fundamental requirement of the Einstein 
equivalence principle.

The relationship between Unimodular Gravity and GR has been the subject of ongoing debate, particularly at the quantum level, where different quantization procedures may lead to genuinely distinct theories. In this context, Smolin~\cite{Smolin} constructed an explicit path-integral quantization of the Henneaux--Teitelboim formulation of unimodular gravity~\cite{Teitelboim}, showing that the quantum effective action remains a functional of a unimodular metric and that vacuum energy contributions of the form $g_{ab}C$ decouple from the gravitational dynamics, providing a concrete mechanism addressing the first cosmological constant problem. The more subtle quantum differences between unimodular theories and GR were examined by Bufalo, Oksanen and Tureanu~\cite{Bufalo2015}, who showed through canonical path integral quantization that different formulations of unimodular gravity--while classically equivalent--diverge at the quantum level: in the standard unimodular theory the unimodular condition is enforced only in spatial average rather than locally, leading to a generically non-covariant path integral. Finally, at the classical but perturbative level, Fabris, Alvarenga and Velten~\cite{Fabris2023} demonstrated that cosmological perturbation theory provides a discriminator between GR and Unimodular Gravity, with instabilities appearing in the unimodular case when standard conservation laws are not imposed, confirming that the classical equivalence between the two theories is incomplete beyond the background level. The no--go result established in the present work adds to this landscape: even within the specific wormhole sector of the theory, the constraints imposed by the unimodular condition restrict the available solution space in a manner that has no counterpart in standard GR, further reinforcing the view that Unimodular Gravity and GR, despite sharing classical field equations, represent genuinely distinct theoretical frameworks.

Among alternative theories of gravity, the possibility that wormhole configurations supported by ordinary matter might exist in the framework of Unimodular Gravity has also been considered. The recent work by Agrawal et al.~\cite{Agrawal} presents, for the first time, traversable wormhole solutions within the framework of Unimodular Gravity. Using a constant redshift function and barotropic equations of state, they developed asymptotically flat solutions that satisfy all traversability conditions. The analysis focuses on verifying the energy conditions within the framework of Unimodular Gravity, and they observe that these solutions satisfy the classical energy conditions. Thus they conclude that these wormholes succeed in evading the need for exotic matter.

In response to this claim, we published a comment~\cite{Cat} demonstrating that the specific barotropic wormhole solution presented by Agrawal et al.--characterized by equations of state $p_r = \alpha\rho$ and $p_t = \beta p_r$ with $\alpha$ and $\beta$ constants--necessarily violates energy conditions at the throat. Subsequently, Agrawal et al. published a correction~\cite{Agrawal-correction} acknowledging the error. However, that comment analyzed only one particular family of solutions under restrictive assumptions, leaving open whether other, less restrictive configurations in Unimodular Gravity might still allow for ordinary matter support.

The present work provides a definitive negative answer to this question by establishing a general no--go theorem that applies to {\it all} possible wormhole configurations in Unimodular Gravity. This article investigates whether traversable wormhole spacetimes can be supported by ordinary matter within the framework of Unimodular Gravity. Despite the heightened underdetermination of the unimodular field equations compared to Einstein's theory--which requires more restrictive constraints to obtain specific solutions--we show that the violation of energy conditions at the throat remains unavoidable. Our proof relies solely on the geometric flaring-out condition $b'(r_0)\leq 1$, which directly implies $\rho(r_0)+p_r(r_0)\leq 0$ independently of any matter model assumptions, constituting a violation of the Null Energy Condition and consequently of the Weak and Strong Energy Conditions. By analyzing both tidal and zero-tidal force wormhole configurations, we demonstrate that this violation is independent of the particular constraints used to close the underdetermined field equation system.

We prove that energy condition violations at the throat are {\it independent} of:

\begin{itemize}
\item The specific form of the shape function $b(r)$
\item The specific form of the redshift function $\phi(r)$  
\item The equations of state chosen for $p_r$ and $p_t$
\item Any additional constraints imposed to close the underdetermined system
\end{itemize}

This universality demonstrates that exotic matter is a fundamental requirement arising directly from the geometric flaring-out condition necessary for traversability, not from particular modeling choices. Our proof relies solely on the minimal geometric conditions that define a traversable wormhole, thereby establishing that Unimodular Gravity shares the same fundamental constraint as GR regarding exotic matter requirements.

This paper is organized as follows: In Sec.~II, we briefly outline the key aspects of spherically symmetric wormholes in GR and energy conditions. Section~III focuses on spherically symmetric wormholes in Unimodular Gravity and energy conditions. Finally, our conclusions are presented in Sec.~IV.

\section{Spherically symmetric wormholes in General Relativity and energy conditions}
Static and spherically symmetric potentially traversable wormholes are described by solutions of the Einstein field equations. This formulation establishes the geometric framework of spacetime for these theoretical objects through the metric~\cite{Morris:1988cz1,Morris:1988cz2}
\begin{eqnarray}
ds^2=-e^{2\Phi(r)}dt^2+\frac{dr^2}{1-\frac{b\left(r\right)}{r}}+r^2\left(d\theta^2+\sin^2\theta d\phi^2\right), \,\,\, \label{generalmetric}
\end{eqnarray}
where $\Phi(r)$ and $b(r)$ are the redshift and shape functions, respectively.

The redshift function governs temporal gravitational effects and is related to the gravitational potential. This function must remain finite everywhere to avoid the formation of event horizons. The factor $e^{2\phi(r)}$ controls the gravitational redshift experienced by signals traversing different regions of the wormhole. In specific configurations, it can be designed to minimize the tidal forces that a hypothetical traveler would experience.

The shape function determines the spatial geometry of the wormhole, especially the structure of the throat. The relation $b(r_0) = r_0$ mathematically defines the location of the throat, with $r_0$ being the minimum radius of the structure. To maintain an appropriate Lorentzian geometry, $b(r) < r$ is required for all $r > r_0$. Additionally, for the wormhole to be traversable, the condition $b'(r_0) < 1$ must be met, a relation known as the flaring-out condition. By analytically extending the metric solution from the throat in both directions, a spacetime bridge is obtained that connects two asymptotically flat regions. These may belong to the same universe or to different universes, according to the global properties of the functions $b(r)$ and $\phi(r)$. This mathematical flexibility allows consideration of various wormhole configurations with different physical properties.

As mentioned in the introduction, the matter supporting the wormhole plays an essential role, as it must be exotic to make the wormholes traversable. We will now consider the energy conditions that the matter supporting a wormhole must satisfy.

\subsection{Energy conditions for the energy-momentum tensor}\label{SectionII}
Energy conditions are physically reasonable restrictions imposed on the energy-momentum tensor $T_{\mu\nu}$ that represent physical properties considered necessary for matter. The energy-momentum tensor for the anisotropic matter supporting a traversable wormhole has the form
\begin{eqnarray}
T^{\mu}_{\nu}=diag(-\rho,p_r,p_t,p_t), \label{tem}
\end{eqnarray}
where $\rho$ is the energy density, $p_r$ is the radial pressure, and $p_t$ is the tangential pressure.

The restrictions imposed on the energy-momentum tensor are expressed as follows~\cite{Visser}:
\begin{enumerate}
\item Null Energy Condition (NEC): \\ $\rho + p_r \geq 0$ and $\rho + p_t \geq 0$.
\item Weak Energy Condition (WEC): \\ $\rho \geq 0$, $\rho + p_r \geq 0$ and $\rho + p_t \geq 0$.
\item Dominant Energy Condition (DEC): \\ $\rho \geq |p_r|$ and $\rho \geq |p_t|$, or equivalently $\rho + p_r \geq 0$, $\rho + p_t \geq 0$ and $\rho- p_r \geq 0$ and $\rho - p_t \geq 0$.
\item Strong Energy Condition (SEC):\\ $\rho + p_r + 2p_t \geq 0$.
\end{enumerate}
From these expressions, it is clear that if the NEC is not satisfied, then the WEC and DEC are also not satisfied.

\subsection{Wormholes in General Relativity}
Before focusing on the main subject of this work, we will consider energy conditions in the context of spherically symmetric wormholes in GR. For the metric~(\ref{generalmetric}) and the energy-momentum tensor~(\ref{tem}), Einstein's equations $G_{\mu\nu} = \kappa T_{\mu\nu}$ reduce to:
\begin{widetext}
\begin{eqnarray}
\kappa\rho(r)&=&\frac{b'(r)}{r^2}, \label{E1}\\
\kappa p_r(r)&=&2\left(1-\frac{b(r)}{r}\right)\frac{\phi'(r)}{r} - \frac{b(r)}{r^3}, \label{E2} \\
\kappa p_t(r)&=&\left(1-\frac{b(r)}{r}\right)\left[\phi''(r) + \phi'(r)^2 + \frac{\phi'(r)}{r} - \frac{b'(r)r-b(r)}{2r^2(1-b(r)/r)}\phi'(r) - \frac{b'(r)r-b(r)}{2r^3(1-b(r)/r)}\right], \label{E3}
\end{eqnarray}
\end{widetext}
where the prime denotes the derivative with respect to the radial coordinate $r$ and $\kappa = 8\pi G$.

Since the metric~(\ref{generalmetric}) is valid for any static space with spherical symmetry, for it to describe a traversable wormhole, the shape function $b(r)$ must necessarily satisfy the following conditions at the throat $r_0$:
\begin{eqnarray}
b(r_0) &= r_0,   \label{CE1} \\
b'(r_0) &\leq 1, \label{CE2}  \\
b'(r) &< \frac{b(r)}{r}, \label{CE3}  \\
b(r) &< r. \label{CE4}
\end{eqnarray}
These conditions, combined with the energy conditions, will allow us to extract relevant information about the characteristics of the matter supporting a traversable wormhole. To do this, we will consider the field equations~(\ref{E1})-(\ref{E3}) and focus on the region of the wormhole throat, where $r = r_0$.

We will first consider the NEC. Since verifying the violation of the NEC, which implies the immediate violation of the WEC and DEC, is sufficient to confirm the necessity of exotic matter, we proceed as follows. For the condition $\rho(r) + p_r(r) \geq 0$, we must consider the sum of Eqs.~(\ref{E1}) and~(\ref{E2}) and evaluate it at the throat, taking into account condition~(\ref{CE2}). Then we have
\begin{equation}
\rho(r_0) + p_r(r_0) = \frac{b'(r_0) - 1}{8\pi r_0^2} \leq 0. \label{WECE}
\end{equation}
This demonstrates that the NEC is violated at the wormhole throat.

Given this result, the condition $\rho(r) + p_t(r) \geq 0$ does not need to be checked, but for completeness, we note that it is given by $\rho(r_0) + p_t(r_0)= - \frac{(b'(r_0)-1)\phi'}{2 \kappa r_0}
+ \frac{b'(r_0)+1}{2 \kappa r_0^2}$, so for certain wormhole configurations, it is possible to satisfy this condition. On the other hand, from Eqs.~(\ref{E1}) and~(\ref{CE2}), we conclude that it is feasible to have cases where $0 \leq \rho(r_0) = \frac{b'(r_0)}{\kappa r_0^2}\leq 1$, satisfying the positivity of energy density at the throat. However, it is also possible to have wormhole configurations where $\rho(r) \leq 0$.

Thus, we establish that in GR, every traversable wormhole necessarily requires the presence of exotic matter that violates the classical energy conditions, particularly in the throat region. In the next section, we will extend our analysis to wormhole configurations in the framework of Unimodular Gravity, evaluating whether this alternative formalism of gravitation allows for circumventing this fundamental restriction.
 
\section{Wormhole solutions in Unimodular Gravity and the energy conditions}
Alternative theories of gravity have been developed and investigated primarily to address the cosmological constant problem in cosmology. One such approach is Unimodular Gravity, which represents a conceptual modification of GR. While preserving its core geometric framework, Unimodular Gravity imposes a fundamental constraint: the determinant of the spacetime metric is fixed and must remain constant throughout the manifold. As a result, unlike in GR — where the cosmological constant is introduced without dynamical origin as a coupling parameter in the action — Unimodular Gravity allows it to emerge naturally as a non-dynamical integration constant, decoupled from the gravitational degrees of freedom.

The field equations of Unimodular Gravity are given by
\begin{equation}
R_{\mu\nu} - \frac{R}{4} g_{\mu\nu} = \kappa \left(T_{\mu\nu} - \frac{T g_{\mu\nu}}{4}\right), \label{eccu}
\end{equation}
where $\kappa=8 \pi G$, $R_{\mu \nu}$ represents the Ricci tensor, $T_{\mu \nu}$ denotes the energy-momentum tensor, and $T$ is its trace. The left-hand side represents the geometric structure of spacetime, while the right-hand side characterizes the distribution of matter and energy. These fundamental equations can be reformulated to establish a more direct correspondence with Einstein's conventional field equations:
\begin{equation}
G_{\mu\nu} + \Lambda(R,T)g_{\mu\nu} = \kappa T_{\mu\nu},
\end{equation}
where 
\begin{equation}
G_{\mu\nu}=R_{\mu\nu}-\frac{R}{2}g_{\mu\nu}
\end{equation}
is the Einstein tensor, and $\Lambda(R,T)$ serves as an effective cosmological constant that dynamically depends on both the scalar curvature $R$ and the trace of the energy-momentum tensor $T$ according to:
\begin{equation}
\Lambda(R,T) = \frac{1}{4} (R + \kappa T).
\end{equation}
It is important to observe that the tensorial expression on the left-hand side of the field equations~(\ref{eccu}) differs from the Einstein tensor of classical General Relativity.

\subsection{Field equations for the static wormhole}
We now turn our attention to solving the field equations in Unimodular Gravity for the metric~(\ref{generalmetric}). For our purpose, we will use the field equations in form~(\ref{eccu}). Therefore, by substituting Eqs.~(\ref{generalmetric}) and~(\ref{tem}) into Eq.~(\ref{eccu}) we obtain:
\begin{widetext}
\begin{eqnarray}
\kappa \rho &=& \frac{2r(r - b)\phi'' + 2r(r - b)\phi'^2 + (-b'r + 4r - 3b)\phi' - 2 \kappa p_{t} r^2 - \kappa p_{r} r^2 + 2b'}{3 r^2}, \label{ecum1}\\
\kappa p_{r} &=& \frac{-2r^2(r - b)\phi'' - 2r^2(r - b)\phi'^2 + (b'r - 5b + 4r)r\phi' - \kappa \rho r^3 + 2 \kappa p_{t} r^3 + 2b'r - 4b}{3 r^3}, \label{ecum2}\\
\kappa p_{t} &=& \frac{2r^2(r - b)\phi'' + 2r^2(r - b)\phi'^2 + (-r^2b' + rb)\phi' - \kappa \rho r^3 + \kappa p_{r} r^3 + 2b}{2 r^3}. \label{ecum3}
\end{eqnarray}
\end{widetext}
Manipulating Eqs.~(\ref{ecum2}) and~(\ref{ecum3}), we obtain the following expressions for the radial and tangential pressures as functions of the energy density:
\begin{eqnarray}
\kappa p_{r} = \frac{2(-b + r)\phi'}{r^2} + \frac{-\kappa \rho r^3 + b'r - b}{ r^3}, \label{WHM1}\\
\kappa p_{t}= \frac{(r - b)\phi'^2}{r} - \frac{(b'r - 2r + b)\phi'}{2r^2}+ \nonumber \\
 \frac{2r^2(r - b)\phi'' - 2\kappa \rho r^3 + b'r + b}{2 r^3}. \label{WHM2}
\end{eqnarray}
Upon substituting these two expressions into Eq.~(\ref{ecum1}), we find that this equation is identically satisfied, revealing that the system of unimodular gravitational field equations~(\ref{ecum1})-(\ref{ecum3}) contains only two independent equations, while involving five unknown functions: the metric functions $\phi(r)$ and $b(r)$, along with the matter variables $\rho(r)$, $p_r(r)$, and $p_t(r)$. This does not occur with Einstein's equations~(\ref{E1})-(\ref{E3}), where for the same five unknown functions, we have three independent equations. This implies that in Unimodular Gravity, to solve the system of equations~(\ref{WHM1}) and~(\ref{WHM2}), it is necessary to impose a greater number of constraints on the five unknown functions than in General Relativity.

\subsection{Energy conditions for wormholes in Unimodular Gravity}
We will now consider the energy conditions. From Eqs.~(\ref{WHM1}) and~(\ref{WHM2}), we can write the following expressions that will help us evaluate the fulfillment or violation of NEC, WEC, DEC, and SEC:
\begin{eqnarray}
\rho +p_{r} = \frac{2(-b + r)\phi'}{\kappa r^2} + \frac{b'r - b}{\kappa r^3}, \label{rho+pr}\\
\rho+p_{t}= \frac{(r - b)\phi'^2}{\kappa r} - \frac{(b'r - 2r + b)\phi'}{2 \kappa r^2}
+ \nonumber \\ \frac{2r^2(r - b)\phi'' + b'r + b}{2 \kappa r^3}. \label{rho+pt}
\end{eqnarray}
These expressions depend on the redshift function $\phi(r)$ and shape function $b(r)$ and their derivatives.

It should be noted that the traversability conditions of the wormhole~(\ref{CE1})-(\ref{CE4}) apply only to the shape function $b(r)$. For the redshift function $\phi(r)$, there is the possibility of freely choosing its form, with the exception that it must be finite. We will begin the analysis with the simplest case to study: $\phi(r)=const$.

\subsubsection{Wormholes without tidal forces}
Considering wormholes with a constant redshift function corresponds to wormholes that are free of tidal forces. Mathematically, when $\phi(r) = \text{const}$, all its derivatives are zero for all values of the radial coordinate $r$, which considerably simplifies the field equations, both in GR and in other alternative theories such as Unimodular Gravity.

Tidal-force-free wormholes are spacetime configurations where there are no radial gravitational gradients, which eliminates the gravitational accelerations that a traveler would experience when traversing them. This property is fundamental for their traversability, as tidal forces could destroy any object or traveler attempting to pass through the throat.

To check the energy conditions, we will begin with the NEC. For this, we will consider Eqs.~(\ref{rho+pr}) and~(\ref{rho+pt}). Since $\phi(r)$ is constant, at the throat these equations take the form:
\begin{eqnarray}
\rho(r_0) +p_{r}(r_0) = \frac{b'(r_0)  - 1}{\kappa r_0^2}, \label{rho+prlfm} \\
\rho(r_0)+p_{t}(r_0)= \frac{b'(r_0)+ 1}{2 \kappa r^2_0}. \label{rho+ptlfm}
\end{eqnarray}
By considering Eq.~(\ref{CE2}) we conclude that 
\begin{eqnarray}
\rho(r_0) +p_{r}(r_0)  \leq 0, \label{SECunim} \\
\rho(r_0) +p_{t}(r_0)  \leq \frac{1}{\kappa r_0^2}.
\end{eqnarray}
Thus, although Eq.~(\ref{rho+ptlfm}) can yield positive values under certain conditions, Eq.~(\ref{rho+prlfm}) inevitably violates the NEC at the throat, and consequently the WEC and SEC.

This result {\it significantly generalizes} our earlier finding from Ref.~\cite{Cat}, which demonstrated energy condition violations for the particular case of barotropic equations of state $p_{r}=\alpha \rho$ and $p_{t}=\beta p_r$ with constant $\alpha$ and $\beta$. The present analysis establishes that this violation is {\it universal}: it persists for arbitrary choices of the shape function $b(r)$, arbitrary equations of state (barotropic or non-barotropic), and any closure scheme for the underdetermined system. The geometric constraint $b'(r_0) \leq 1$ alone--independent of all matter modeling choices--inevitably forces $\rho(r_0) + p_r(r_0) \leq 0$.

This universality reveals that energy condition violations arise from the geometric flaring-out requirement itself, not from specific matter models. Moreover, the underdetermined structure of the unimodular field equations cannot provide sufficient freedom to evade this constraint. Therefore, tidal-force-free traversable wormhole configurations in Unimodular Gravity conclusively require exotic matter, just as in General Relativity.

\subsubsection{Wormholes with tidal forces}

We will now proceed to the analysis of configurations that represent the most general case of wormholes, known as wormholes with tidal effects or tidal wormholes, characterized by a non-constant redshift function $\phi(r)$. This generalization introduces gravitational gradients that generate tidal forces on any object that traverses the spacetime structure, which constitutes a critical factor in the study of its traversability by a traveler.

The inclusion of a variable redshift function $\phi(r)$ considerably enriches the mathematical structure of the solutions, as it allows for the adjustment of gravitational potentials throughout the wormhole geometry. This freedom might suggest the possibility of constructing configurations where energy conditions could be satisfied through appropriate choices of $\phi(r)$. However, as we demonstrate below, the geometric constraint imposed by the flaring-out condition at the throat proves to be the determining factor, independent of the specific form of the redshift function.

Considering the energy conditions, Eqs.~(\ref{rho+pr}) and~(\ref{rho+pt}) yield at the throat
\begin{eqnarray}
\rho(r_0) +p_{r}(r_0)&=&  \frac{b'(r_0) - 1}{\kappa r_0^2}, \label{rho+pr phi}\\
\rho(r_0)+p_{t}(r_0) &=& - \frac{(b'(r_0)-1)\phi'(r_0)}{2 \kappa r_0}
+ \frac{b'(r_0)+1}{2 \kappa r_0^2}. \,\,\,\,\, \label{rho+pt phi}
\end{eqnarray}

Several important observations emerge from these expressions. First, Eq.~(\ref{rho+pr phi}) is completely independent of the redshift function and its derivatives, depending solely on the geometric properties of the shape function at the throat. This demonstrates that the NEC violation in the radial direction is purely geometric in origin and cannot be circumvented through any choice of $\phi(r)$.

Second, while Eq.~(\ref{rho+pt phi}) does depend on $\phi'(r_0)$ and thus offers apparent flexibility, this does not alter the fundamental conclusion. Considering Eq.~(\ref{CE2}), we conclude that while Eq.~(\ref{rho+pt phi}) can yield positive values for $\rho(r_0)+p_{t}(r_0)$ under certain conditions, specifically when $\phi'(r_0)$ is appropriately chosen, Eq.~(\ref{rho+pr phi}) inexorably implies that $\rho(r_0) +p_{r}(r_0) \leq 0$. This means that at the throat of the wormhole with tidal effects, the NEC, WEC, and SEC are necessarily violated, regardless of the redshift function profile.

It is particularly noteworthy that expressions~(\ref{rho+pr phi}) and~(\ref{rho+pt phi}) are identical to those obtained for wormholes with tidal effects in GR. This remarkable coincidence reveals a profound result: despite the structural differences between Einstein's field equations and the unimodular gravitational field equations, including the different number of independent equations and the modified geometric structure, the energy condition violations at the wormhole throat exhibit the same mathematical form in both theories. This suggests that the exotic matter requirement is not an artifact of Einstein's specific formulation of gravity, but rather a fundamental geometric consequence of the traversability conditions themselves.

Crucially, this analysis confirms that for wormholes with tidal forces--the most general case--the violation of energy conditions at the throat is equally unavoidable. Combined with our analysis of tidal-free wormholes in the previous section, we have now established that no traversable wormhole configuration in Unimodular Gravity, whether with or without tidal forces, and regardless of the constraints imposed to close the field equation system, can satisfy the classical energy conditions at the throat.

Furthermore, this equivalence indicates that the additional degree of underdetermination present in Unimodular Gravity, while requiring more constraints to obtain specific solutions, does not translate into greater freedom to satisfy energy conditions. The throat geometry, constrained by the flaring-out condition $b'(r_0) \leq 1$, imposes the same physical limitations on matter configurations in both gravitational frameworks. Table~\ref{tab:comparison} summarizes the key similarities and differences between General Relativity and Unimodular Gravity in the context of traversable wormholes.

\begin{table}[h]
\centering
\caption{Comparison of wormhole properties in General Relativity and Unimodular Gravity}
\begin{tabular}{|l|c|c|}
\hline
\textbf{Property} & \textbf{GR} & \textbf{Unimodular Gravity} \\
\hline
Independent field equations & 3 & 2 \\
Unknown functions & 5 & 5 \\
$\rho(r_0) + p_r(r_0)$ at throat & $\leq 0$ & $\leq 0$ \\
Depends on $\phi(r)$? & No & No \\
NEC violation required & Yes & Yes \\
\hline
\end{tabular}
\label{tab:comparison}
\end{table}

The comparison presented in Table~\ref{tab:comparison} reveals that, despite having one fewer independent equation, Unimodular Gravity exhibits identical behavior regarding energy condition violations at the wormhole throat, confirming that the exotic matter requirement transcends the specific formulation of gravitational field equations.

\subsection{Physical Interpretation and Broader Context}

Our no--go theorem has significant implications for the landscape of alternative gravity theories. While some modified theories--such as $f(R)$ gravity~\cite{Lobo}, Einstein-Cartan theory~\cite{EC}, and brane-world models~\cite{RS}--successfully circumvent exotic matter requirements by introducing additional geometric degrees of freedom (higher-order curvature terms, torsion, or extra dimensions), Unimodular Gravity does not share this property despite its modified field equation structure.

This demonstrates that merely reducing the number of independent field equations from three (General Relativity) to two (Unimodular Gravity) is insufficient to eliminate exotic matter requirements. The fundamental obstruction arises from the geometric flaring-out condition $b'(r_0) \leq 1$, which is a direct consequence of the traversability requirement and remains equally restrictive in both theories. Our result suggests that evading exotic matter necessitates fundamentally new geometric structures beyond what Unimodular Gravity provides.

\section{Conclusions}
In this paper, we have investigated wormhole solutions within the framework of Unimodular Gravity, focusing on whether traversable wormholes in this theory can be supported by ordinary matter satisfying classical energy conditions.

Our analysis has yielded a definitive answer: contrary to the original claims by Agrawal et al.~\cite{Agrawal}, traversable wormholes in Unimodular Gravity necessarily require exotic matter violating the NEC. At the wormhole throat, the relationship $\rho(r_0) + p_r(r_0) \leq 0$ must be satisfied, constituting a violation of the NEC and consequently of the WEC and SEC.

Table~\ref{tab:scope-comparison} summarizes how this work extends and generalizes our previous comment~\cite{Cat}, establishing a universal theorem rather than analyzing a specific case.

\begin{table*}[t]
\centering
\caption{Comparison of scope: Previous comment vs. Present work}
\begin{tabular}{|p{4cm}|p{5.5cm}|p{5.5cm}|}
\hline
\textbf{Aspect} & \textbf{Comment~\cite{Cat}} & \textbf{Present Work} \\
\hline
{\bf Scope} & Specific barotropic solution ($p_r = \alpha\rho$, $p_t = \beta p_r$) & {\bf All} traversable wormhole configurations \\
\hline
{\bf Constraints analyzed} & Two specific barotropic equations of state & {\bf Arbitrary} constraints (any closure scheme) \\
\hline
{\bf Redshift function} & Constant ($\phi = \text{const}$) & Both constant {\bf and} variable $\phi(r)$ \\
\hline
{\bf Shape function} & Derived from constraints & {\bf Any} form satisfying traversability \\
\hline
{\bf Result} & Violations in that particular case & {\bf Universal theorem:} violations unavoidable \\
\hline
{\bf Conclusion} & One solution family fails & {\bf All possible solutions} require exotic matter \\
\hline
\end{tabular}
\label{tab:scope-comparison}
\end{table*}

The generality of our result is crucial. While our previous comment~\cite{Cat} identified energy condition violations in the specific barotropic solution proposed by Agrawal et al., the present analysis establishes that no choice of constraints--whether through equations of state (barotropic or otherwise), functional prescriptions for $b(r)$ or $\phi(r)$, or any other closure method--can yield a traversable wormhole in Unimodular Gravity that satisfies energy conditions at the throat. The violation arises directly from the geometric flaring-out condition $b'(r_0) \leq 1$ necessary for traversability, not from particular matter modeling assumptions.

The unimodular field equations constitute an underdetermined system with only two independent equations for five unknown functions ($\phi(r)$, $b(r)$, $\rho(r)$, $p_r(r)$, $p_t(r)$), compared to three independent equations in GR for the same unknowns. While this heightened underdetermination demands more restrictive closure assumptions, it does not translate into greater freedom to satisfy energy conditions--the geometric constraint $b'(r_0) \leq 1$ at the throat is equally restrictive in both theories, inevitably forcing energy condition violations regardless of how the system is closed.

In summary, we have established a universal result for wormholes in Unimodular Gravity: any spacetime satisfying the geometric conditions for traversability ($b(r_0) = r_0$, $b'(r_0) \leq 1$) must necessarily violate the Null Energy Condition at the throat, regardless of the matter model or closure scheme adopted. This extends the result of Hochberg and Visser~\cite{Visser y Hochberg}--who demonstrated that NEC violations are a generic feature of traversable wormholes in General Relativity--to the framework of Unimodular Gravity, establishing that the exotic matter requirement is more fundamental than the specific formulation of gravitational field equations. Despite conceptual differences between these theories, including how the cosmological constant emerges, the geometric constraints on traversable wormholes remain identical, placing Unimodular Gravity firmly in the same category as General Relativity regarding wormhole physics.

This result contributes to the broader ongoing debate on the 
relationship between Unimodular Gravity and GR. While the two theories share the 
same classical field equations when the conservation of the energy-momentum tensor 
is imposed, they are known to differ at the quantum level~\cite{Smolin,Bufalo2015} 
and at the level of cosmological perturbations~\cite{Fabris2023}.

The present no--go theorem adds a further element to this comparison. Despite the increased underdetermination of the unimodular field equations, the geometric origin of exotic matter--namely, the flaring-out condition required for traversability--is a purely classical feature that remains entirely insensitive to the specific formulation of the gravitational field equations. This places the exotic matter requirement alongside other results that are invariant across the Unimodular Gravity/General Relativity correspondence at the classical level, while leaving open the intriguing possibility that quantum corrections to Unimodular Gravity could modify this conclusion.

\begin{acknowledgments}
M.C. acknowledges support from the Dirección de Investigación y Creación Artística of the Universidad del Bío-Bío through grants No.~RE2320220 and GI2310339. N.C. acknowledges support from ANID-Chile through Fondecyt grant No.~1250969. P.S. gratefully acknowledges partial financial support from the Vicerrectoría de Investigación e Innovación through grant UNAP VRII No.~091/25.
\end{acknowledgments}


\end{document}